# Controlled defects in ZnO by low energy Ar irradiation

Soubhik Chattopadhyay[1], Sreetama Dutta[1], D. Jana[1*], S. Chattopadhyay[2], A. Sarkar[3], P. Kumar[4], D. Kanjilal[4], D. K. Mishra[5], S. K. Ray[6]

[1]Department of Physics, University of Calcutta, 92 Acharya Prafulla Chandra Road, Kolkata-700 009, India.
[2]Department of Physics, Taki Government College, Taki -743429, India.
[3]Department of Physics, Bangabasi Morning College, 19 Rajkumar Chakraborty Sarani, Kolkata -700 009, India.
[4]Inter-University Accelerator Centre, P.O. Box 10502, Aruna Asaf Ali Marg, New Delhi 110067, India.
[5]Advanced Materials Technology Department, Institute of Minerals and Minerals Technology, Bhubaneswar – 751013, India
[6]Department of Physics and Materials Science, Indian Institute of Technology, Kharagpur, India.

---

* Author to whom correspondence should be addressed; electronic mail: djphy@caluniv.ac.in

**Abstract**

We report interesting observations in 1.2 MeV $Ar^{8+}$ ion irradiated ZnO which, to the best of our knowledge, have not been published earlier and will be useful for the scientific community engaged in research on ZnO. Possible amorphisation in ZnO has not been noticed even in the highest irradiation fluence $5 \times 10^{16}$ ions/cm$^2$. Irradiation with the initial fluence $1 \times 10^{15}$ ions/cm$^2$ changes the colour of the sample from white to orange while the highest irradiation fluence makes it dark reddish brown that appears as black. Such changes in colour can be correlated with the oxygen vacancy type defects. No significant change in the grain size of the irradiated samples, as revealed from the x-ray diffraction (XRD) line width broadening, has been observed. However, with initial irradiation fluence relaxation of strain and/or little recovery of defects have taken place. Increase of surface roughness due to sputtering is clearly visible in scanning electron micrographs (SEM) with highest fluence of irradiation. Room temperature Photoluminescence (PL) spectrum of the unirradiated sample shows intense ultra-violet (UV) emission (~ 3.27 eV) and less prominent defect level emissions (2-3 eV). The overall emission is largely quenched due to initial irradiation fluence. But with increasing fluence UV emission is enhanced along with prominent defect level emissions. This is contrary to the observed features from XRD. Very small increase of XRD line width have been observed for highest fluence. Remarkably, the resistivity of the irradiated sample with highest fluence is reduced by four orders of magnitude compared to that of the unirradiated sample. This indicates increase of donor concentration as well as their mobility due to irradiation. Oxygen vacancies are deep donors in ZnO, but surely they influence the stability of the shallow donors (presumably zinc interstitial related) and vice versa. This is in conformity with recent theoretical calculations.

Defects in ZnO and their role on structural, electronic and optical properties are being widely discussed at present. The extreme complexity of defects in ZnO has added merit to this agenda[1-7]. In one way, defect-induced tuning of material properties according to the technological need is becoming more and more prospective. For example, fabrication of p-type ZnO[8] (in general, ZnO bears intrinsic n-type conductivity) or developing UV-visible luminescent and lasing device with ZnO[9] or realizing ferromagnetic behavior in diamagnetic ZnO lattice[10] solely depends on the purposeful defect management in the material. On the other side, efforts to understand and alter the defective state in ZnO are providing new and newer insights on the electronic and optical processes in condensed matter. This aspect of ZnO research is no less challenging than fabrication of devices. New theoretical approaches are being developed[1,5,6,11-13] which have generated enormous interest and also expanded the horizon of theoretical condensed matter physics.

Therefore, controlled defect production in ZnO, defect characterization by suitable techniques and understanding the defect-property correlation is to be prime focus as of now. Controlled incorporation of defects generally involves high temperature annealing[14], substitution[15], mechanical milling[16] and irradiation by energetic particles[4,17]. With the first three techniques, one can monitor the equilibrium defect configuration (from requirement of the free energy minimization) in the system. Energetic ion beam irradiation is an efficient way to induce non-equilibrium defect states in solid materials[18]. In this paper, we have studied the effect of 1.2 MeV $Ar^{8+}$ ion irradiation on polycrystalline ZnO (Purity 99.99%, Sigma-Aldrich, Germany) from XRD, SEM, room temperature resistivity and photoluminescence (PL) measurements. SEM reveals the surface morphology of the samples whereas PL characteristics bear signature of few tens of nanometers (taking absorption coefficient $1.6 \times 10^5$ cm$^{-1}$ at 325 nm[3,19]) below the surface. XRD penetrates (1.54 Å Cu K$_\alpha$ radiation) down to ~ 2 μm[19] in ZnO while resistivity is purely a bulk property. All the prepared ZnO pallets are annealed in air at 500 °C. The pre-annealing before irradiation has been carried out to make the sample free from residual organic materials[10,20] or $H_2$[21], if any. The annealed samples are cooled slowly at a rate of 36 °C/h down to room temperature to reduce unwanted defect (mostly interstitial type) freezing during cooling[2,22,23]. Ar ion has been chosen to minimize possible chemical or magnetic effects in the target. Moreover, Ar has an atomic mass intermediate to Zn and O and it is interesting to observe relative effects on Zn and O sites by its impact. Recently, Wendler et al. have investigated[24] the effect of 200 KeV Ar beam on ZnO single crystal and have observed a saturation of disorder above fluence $10^{16}$ ions/cm$^2$ ( ~ 8 displacements per atom). Borges et al. have reported[25] the occurrence of ferromagnetism at room temperature due to defects generated by 100 KeV Ar ion in single crystalline ZnO. Krishna et al. have studied[26] the modification of optical properties in nanocrystalline ZnO thin films by 15 KeV Ar beam. Kucheyev et al. have mentioned[17] heavy damage in ZnO lattice by low KeV Ar ion without detailing it. Nevertheless, the understanding on the ion induced defects in ZnO is rather incomplete. Particularly, reports on the modification of photoluminescence properties of ZnO by low energy ion irradiation are very much limited and not comprehensive too.

As-supplied ZnO samples have been annealed in air at 500 °C for 4 hours in a programmable digital furnace and subsequently irradiated with 1.2 MeV Ar$^{+8}$ ions at fluences $1 \times 10^{15}$, $5 \times 10^{15}$ and $5 \times 10^{16}$ ions/cm$^2$ using low energy ion beam facility (LIBEF) at Inter University Accelerator Centre (IUAC), New Delhi. XRD studies have been carried out using Philips PW 1830 automatic powder diffractometer with Cu $K_\alpha$ radiation. The range of scanning is 10°- 80° (2θ) in steps of 0.01°. SEM has been recorded in SEM-S3400N, Hitachi, Japan. Room temperature (RT) PL measurement has been carried out with He-Cd laser as an excitation source, operating at 325 nm with an output power 45 mW and a TRIAX 320 monochromator fitted with a cooled Hamamatsu R928 photomultiplier detector.

The impinging charged projectile (here $Ar^{8+}$) in the target suffers both elastic and inelastic collisions. The inelastic collision contributes the major part of the energy transfer when the velocity of the impinging ion is much greater than the orbital velocity of the K-shell electron(s) bound to the target atom(s). This type of energy transfer (loss) leads to excitation and ionization of target atoms and is known as electronic energy loss ($S_e$). As the ion moves deeper inside the target its velocity becomes much lower than the Bohr velocity (0.22 cm/ns). In this regime another factor contributing to its energy loss becomes prominent. The ion suffers series of elastic collisions (with target atom nuclei) near the end of its trajectory. This energy loss of the ion due to elastic collision is known as nuclear energy loss ($S_n$). The total energy loss is the sum of electronic and the nuclear energy losses. The relative contribution of $S_e$ and $S_n$ depends on the projectile mass, velocity, charge state and on the target itself[27]. In this present work, $S_e$ and $S_n$ have been estimated from Stopping Power and Ranges of Ion in Matter (SRIM)[27]. In this numerical computation the density parameter of ZnO has been fixed to 4 gm/cm$^3$. The displacement threshold energy has been taken to be 18.5 eV and 41.4 eV for Zn and O atom in ZnO lattice[4] respectively. The result of the simulation is shown in Fig. 1. The penetration depth of 1.2 MeV Ar is ~ 1 μm compared to the sample thickness ~ 300 μm. $S_e$ largely predominates over $S_n$ except last 300 nm of the ion trajectory. Also, within first 100 nm most of the ion energy is used to excite and ionize target atoms (Fig. 2). It is to be noted here that energy transfer through $S_n$ can only knock out the target atoms from their lattice positions to create stable vacancy/vacancy clusters. $S_e$ is mostly used for exciting the target atom electrons. Above certain critical value of $S_e$ (~ keV/nm), point defects or correlated defect clusters are produced in insulators[28]. Here, the $S_e$ for Ar beam on ZnO is far less than such critical values. SRIM calculation also reveals that Ar beam creates more damage at the Zn site than O sites. However, SRIM only predicts the generated displacements. Majority of such defects immediately face annihilation (dynamic recovery) and that is the origin of radiation hardness[4,17] of ZnO. The actual number of stable defects (after immediate recovery) due to electron irradiation is two orders of magnitude less than the generated defects at the Zn sites[29,30]. The Zn interstitials ($Zn_I$) are mobile even at room temperature with low migration barrier (0.55 eV)[1]. This lowers the number of vacancy and interstitial defects at the Zn sites. The stable oxygen vacancies ($V_O$) (for electron irradiated ZnO) have been estimated[30] to be one order of magnitude higher than the zinc vacancies.

Radiation induced colouration in ZnO is an important area of discussion for decades[31]. Generally, it is believed that irradiation generates anion vacancies (here oxygen vacancies, $V_O$) which creates hydrogen-like bound state with the nearby electrons. This type of defects is commonly known as F-centre or colour centre. The excitations/transitions of this hydrogen like bound state lie in the visible range. However, Vehse et al.[31] have not ruled out the possibility of the occurrence of colour due to $Zn_I$.

Recently, it has been confirmed by Selim et al.[32] and Halliburton et al.[33] that red colouration in ZnO is due to $V_O$s. This is further evident from the fact that the red coloured ZnO becomes colourless transparent by annealing in oxygen atmosphere[32,34]. In the present case, the white samples turns out orange and finally dark reddish brown (appears as black) with increase of irradiation fluence (Fig. 3). It is understood that absorption in the blue green region[33] gives rise to such coloration in ZnO. We have earlier irradiated polycrystalline ZnO by 100 MeV Si, but no significant change in colour has been observed[7]. However, 1 MeV electron irradiation produces red colouration in ZnO[35]. Only after ball milling for a long time gives rise to yellowish ZnO with an absorption edge lowering ~ 0.1 eV[7,36]. Annealing the ball milled material at 1100 °C changes the colour of the sample to reddish. Positron annihilation spectroscopic studies have shown that there is no relation between positron annihilation parameters and the occurrence of reddish colour[16,32]. It is further confirmed by the fact that 1 MeV electron irradiation cannot produce any detectable change in positron annihilation parameters[35]. Calculations show that the threshold energy for stable damages to be produced in the Zn site is ~ 1.6 MeV[4]. Positron annihilation parameters are particularly sensitive for the vacancies or vacancy complexes related to Zn sites and insensitive to isolated $V_O$ at room temperature[7,30]. So, combining all the results mentioned above, it can be concluded at this stage that $V_O$s are responsible for the orange-red colouration of ZnO. With sufficient number of $V_O$s it turns to be dark reddish brown that appears as black. However, a recent investigation shows[37] the appearance of black colour in ZnO with laser irradiation. It has been observed[37] that segregation of metallic zinc has taken place in the irradiated region. In fact, Kucheyev et al. have found[17] preferential loss of oxygen from the subsurface region due to irradiation. Besides, SEM of the sample (Fig. 4) irradiated with highest fluence show signature of sputtering from surfaces. That is why we have measured the resistivity of the irradiated region (for highest fluence) taking the thickness to be 1 μm. Remarkably the resistivity is lowered by four orders of magnitude compared to the unirradiated part. Such a lowering of resistivity have been earlier observed in 100 KeV Ni and 50-380 KeV P irradiated ZnO[38,39]. At the same time occurrence of $V_O$ due to Phosphorous ion irradiation has been identified by Raman spectroscopy[39]. Similar phenomenon of drastic resistance loss (14 orders of magnitude) has been reported due to 175 KeV Li induced native defects in MgO[40]. In MgO, $V_O$s are thought to be responsible for such a huge reduction in resistivity[40]. In this present study, the sample with highest irradiation fluence remains semiconducting down to 77 K. So, the possibility of zinc segregation, at least in our irradiated samples, can be excluded. The possible reason for resisitivity reduction has been discussed later.

The x-ray diffractogram of the unirradiated and irradiated samples show no traces of amorphisation or Zn segregation within the detection limit of the XRD (Fig. 5). This is consistent with the majority of earlier reports[17,38]. However, it has been reported[41] by Vijayakumar et al. that 100 KeV oxygen at fluence $2 \times 10^{16}$ ions/cm$^2$ can induce amorphisation in ZnO. Possible chemical effect of oxygen[41] may be responsible for amorphising ZnO lattice. In fact, 60 KeV Si induces[17] a strong chemical effect in ZnO leading to locally amorphised lattice. This contention has been confirmed[7] from the fact that 100 MeV Si just creates little XRD peak broadening in ZnO. 100 MeV Si penetrates (~ 27 μm) to such a depth which is beyond the penetration depth of 1.54 Å x-rays (~ 2 μm)[19]. So, here, only defects due to irradiation are being probed by XRD and no sign of amorphisation has been observed[7]. As far as XRD results are concerned, in the present work, high fluence ($5 \times 10^{16}$ ions/cm$^2$) of 1.2 MeV Ar ions (penetrating less than the probing x-rays) does not produce any detectable chemical effect. Here, the initial fluence ($1 \times 10^{15}$ ions/cm$^2$) produces two apparently opposite effects in XRD. The FWHM of the (101) peak is lowered as well as the peak intensity (Fig. 6). Such an effect has also been found[38] with 100 KeV Ni beam irradiation. Simultaneous contribution of three major factors can result rich variety of defects in polycrystalline ZnO. The first one is the high dynamic defect recovery of radiation defects in ZnO. The second one originates from its granular nature. It is well known that polycrystalline samples bear a large volume fraction of the material with high concentration of defects (mostly charged), namely grain boundaries (GB)[22]. The grain interiors bear relatively much lower concentration of defects. Effect of irradiation in such co-existent defective and less-defective structure is an interesting topic of discussion[42] for granular materials in general. In high temperature superconductors[42] energetic ion beam degrades the GBs crystal structure to a higher extent with respect the bulk material. Here, the effect appears to be opposite. The high energy loss due to ionization possibly re-organizes the GB region in ZnO. When the region relaxes within few ps, it stabilizes with a fraction of its charged defects recovered[26]. But in the region with relatively lower defects (grain interiors), such re-organization of defects is less efficient. In fact, generation, recovery and stabilization of defects due to some external perturbation should depend strongly on the local defective state of a material if defect-defect interaction is high enough[5]. So, different effects of energetic ion beam at grain boundaries and interiors are possible. The third contribution comes from the nuclear energy loss ($S_n$) related knock out of target atoms. Some stable defects or defect complexes are generated in the whole material[4,17]. Altogether, the granular material becomes more homogenized with defects with much lower volume fraction of segregated defect clusters (grain boundaries) but with scattered defects/defect complexes or their small size clusters. To summarize, the final macroscopic and microscopic defective state (nature of isolated defect clusters and their spatial separation) in ZnO depends on relative loss of projectile energy by inelastic and elastic collisions, $S_e$ and $S_n$ respectively, both of which are functions of ion mass, energy and charge state to some extent. Also the defective state will continuously change as the projectile moves deeper inside the material. The XRD peak intensity and FWHM carry information of the overall defective nature (actually weighted average of contributions from the defective and less-defective regions) within the penetration depth of the x-ray (here ~ 2 μm). If the defect concentration is so high to promote more incoherent scattering, then the XRD peak broadens compared to that of the unirradiated sample[7]. With sufficiently high disorder, destruction of long range lattice order may lead to amorphisation with no XRD peak[41]. If the size of the defect clusters (isolated) is not very large but they are closely spaced then both FWHM as well as peak intensity may reduce[38] as is seen in this study. In case of such defect clusters placed in a relatively higher distance, peak FWHM may decrease with increase of coherent scattering leading to higher peak intensity[43]. A homogenized defect structure evolved in a granular material due to low irradiation fluence should, in principle, lower the residual microstrian in the material. Radiation induced lowering of FWHM due to strain release have been found[44] by Agarwal et al. with 100 MeV Ag ions. With increasing Ar fluence to the $5 \times 10^{15}$ ions/cm$^2$ peak FWHM gets little increment and shows saturation type behavior with further increase of fluence. Such saturation of radiation induced disorder in ZnO has earlier been reported for electron irradiation[45] and Ar irradiation[24]. Look et al. predicted[4] that most abundant stable defect after irradiation is Zn vacancy ($V_{Zn}$)- Zn antisite ($Zn_O$)-Oxygen antisite ($O_{Zn}$) type and remaining $Zn_I$ is placed at a distance to avoid immediate recombination (with $V_{Zn}$) after generation. As mentioned earlier, $Zn_I$s in ZnO are mobile at room temperature. To form stable defect-defect clusters $V_{Zn}$ has to form some complex before reaching $Zn_I$ at that site. Naturally, this condition is valid up to some low critical fluence. In a non-equilibrium condition during high fluence of irradiation, a saturation of defect concentration is more likely because the spatial separation between generated $V_{Zn}$ and $Zn_I$ is close enough to recombine immediately. A competition between defect generation and recovery (ionization induced recovery contributes to some extent) leads to such saturation of defects with high fluence regime[24]. We have also plotted the ratio (R) of peak intensities of (002) and (101) peak (Fig. 7) with fluence. This also illustrates close similarity with the variation of peak FWHM with irradiation fluence (Fig. 7). The intensity of (101) peak in ZnO is believed to indicate[46] the oxygen deficiency to some extent. The variation of R shows dominant defect types at initial fluence and at higher fluence are different. A saturation of defect is also prominent at high fluence regime. The variation of R is not large enough for a conclusive understanding. However, to note, at least 50 % of total region probed by XRD is unaffected by the Ar beam. The variation of the position (2θ) of the (101) peak with fluence also follows the features of the variation of R (Fig. 7). Lowering 2θ is indicative of little expansion of lattice most probably due to increase of $V_O$[47]. At the same time higher 2θ (at initial fluence) indicates that dominant defect types are not $V_O$ but some other type like $V_{Zn}$ related complexes[7].

In the light of the above discussion let us revisit the drastic resistivity loss and the change of colour in the irradiated samples. Theoretically this issue is well discussed[1,6] however, not settled yet. Experimentally, our report is one of the most conclusive reports on this agenda. During thermochemical reduction of ZnO at high temperatures in sealed ampoules[32,33], it is not possible to control the extent of reduction. The final product becomes deep red in colour. A control, however, over the colour of ZnO (orange to red) has been reported through chemical vapour transport technique using Carbone as transport agent[34]. Here also, suitable choice of Ar fluence exhibits gradual but simultaneous evolution of colouration and conductivity in ZnO. Recently this feature has been verified[48] in ZnO single crystal also. The prime reason behind the lowering of resistivity is the modification of GBs due to irradiation (discussed earlier). It is well known that the electrical properties of such granular semiconductors are very much dependent on the defect structure of the GBs[49,50]. That is why we can presume that a huge increase of donor like carriers as well as their mobilities has taken place due to irradiation. Localized carriers at the GB defects become free and the potential barriers at the GB get reduced[50]. This is coherent with the fact that huge resistivity reduction in a similar ZnO sample has been observed[14,16] only after significant grain growth starts due to annealing at elevated temperatures. An intense debate is going on regarding the nature of defects responsible for the supply of donors. Several theoretical models[1] show that oxygen vacancies are deep donors and cannot be the source of conductivity at room temperature. Look et al proposed[4] that $Zn_I$-N related complex might be the dominant donor in ZnO. Lany and Zunger[6] developed a model based on the nonstoichiometry of such oxides and favoured that $V_O$s (doubly ionized) are the source of colour, conductivity as well as persistent phoconductivity in ZnO. Depending on the neighboring cation-cation distance, the vacant oxygen sites can have energy levels close to the conduction band minimum. Their contention has been challenged by Janotti and Van de Walle[1]. The later group suggested hydrogen as the source of donors in ZnO. It has also been reported[51,52] that presence of $V_O$s systematically modify the carrier concentration in ZnO which is not expected due to its deep donor nature. Higher green luminescence has been found[52] for the sample with higher carrier density (which is very close to our results as we shall discuss later). Such a correlation has been wrongly understood earlier as the shallow donor state of $V_O$s. However, Vanheusden et al. have pointed[51] out a systematic difference between free carrier and $V_O$ concentration. Another theoretical calculation[5], published very recently, with defect-defect interaction taken into account predicts a crucial role of $Zn_I$-$V_O$ defect pair in ZnO. Below certain distance these two defects help each other to stabilize. Such stabilization can lower the mobility $Zn_I$. On the other hand the probability of $O_{Zn}$s formation should lead to saturation or even lowering of $V_{Zn}$ during irradiation. Future controlled experiments should focus on the role of $Zn_I$-$V_O$ defect pair. Most probably, $V_{Zn}$-$V_O$ type vacancies near the GBs in polycrystalline ZnO (as predicted and detected in several positron annihilation spectroscopic studies[7]) are dissolved due to irradiation and $Zn_I$-$V_O$ type pairs become dominant (others are $V_{Zn}$, $O_{Zn}$, $Zn_O$ etc.) defects species above certain fluence. Such pairs may help the simultaneous evolution of red color with huge conductivity and UV-defect related PL emissions in polycrystalline ZnO. Finally, we want to note a difference in the effect of annealing and ion irradiation of ZnO. In case of annealing in air, the production of thermally generated $Zn_I$ requires very high temperature at which some Zn vapour also gets released. In such case, lowering of resistivity occurs without any prominent colouration[14]. In our opinion, recovery of defects during such high temperature annealing mostly determines the resistivity lowering. For efficient generation of $Zn_I$, annealing in presence of Zn or Ti vapour in closed ampoules is needed. Generation of $Zn_I$ is also very much efficient with ion irradiation. With high concentration of $Zn_I$ defects, $V_{OS}$ get stabilized (energetically favourable, in other words) to act as colour centers in ZnO.

The room temperature photoluminescence spectra of the unirradiated and irradiated samples have been shown in figure 8. Even after few hundreds of publications[1,3,9,23,51-66] on luminescence in ZnO, consensus among the scientific community on its origin is poor. The unirradiated sample shows UV emission (centered at 3.27 eV) with asymmetric peak shape and FWHM ~ 165 meV. The position and FWHM of the peak is consistent with other reports on polycrystalline ZnO material or thin films[53-55]. A careful observation reveals that the asymmetry of the UV peak is due to the presence of another peak ~ 3.17 eV. The origin of this peak is surely from one of the native defects but exact assignment of the defect is not trivial. Experimental evidences for $Zn_I$[56], $V_{Zn}$[57], dislocation[58] or GB[59] related defects have been reported. Ong et al. have found[60] a peak at 3.13 eV in cathodoluminescence (CL) spectrum which correlates nicely with the Urbach tail parameter (band tail parameter, associated with the defects) of the ZnO film. They have attributed its origin to one of the native defects without specifying it.

The origin of the 3.27 eV peak in the PL spectrum has also been debated. Most of the reports favor the excitonic origin[56] of this peak. However, the room temperature UV peak at such energies in granular ZnO may be an admixture of free exciton as well as some defect related transitions[53,61]. According to Qui et al.[61] such defects energy states are shallow and they reside near the grain boundaries. They assigned this defect giving rise to 3.28 eV peak as $Zn_I$ related defects. However, they favor such recombination is non-excitonic in nature. Fonoberov et al. have investigated[54] the thermal evolution of this UV peak in ZnO nanoparticles (~ 20 nm) with grain size similar to that of ours (~ 35 nm). They attributed the origin of this peak is from donor bound excitons at room temperature.

The origin of defect level emissions (2-3 eV) in ZnO is even more debated[1,3,9,51,62-64]. Actually, such emissions are broad in nature (FWHM ~ 0.4 eV) and so it has contributions from more than one type of defects. Several proposals involving $V_{Zn}^{1,30}$, $Zn_I^{56}$, $V_O^{62}$, $O_{Zn}^{63}$ or impurities[1,3] have been discussed in literature. Efforts continue for an unambiguous identification of the chemical nature of the defect species. Another viewpoint is that the chemical identity of the defects involved is not the relevant issue. Rather, such emissions can originate from disordered part of the lattice[64], more specifically, the disorder at the GBs. At least in the present study, this possibility can be omitted. If the extent of the GB disorder is enhanced due to Ar irradiation, then that should increase the sample resistance. But the sample with highest irradiation fluence shows drastic reduction of resistance along with an increase of emission in the range 2-3 eV.

It is to be noted that the overall emission is largely quenched with irradiation $1 \times 10^{15}$ ions/cm$^2$. This is surely due to appearance of some non-radiative defect centres. Interestingly, XRD peak FWHM is little lowered for this sample. This is due to the different probing region by XRD and PL. XRD probes down to ~ 2 μm below the surface of the sample whereas PL comes from within first 100 nm or so. In the PL scan region ionization induced defect recovery is very high, particularly near the highly defective GBs (discussed earlier). Original GB related defect structure of the unirradiated sample has been heavily modified by irradiation. At the same time $S_n$ induced knock out of the target atoms is also very low in this region. After dynamic recovery of the majority of irradiation induced defects, some defect complexes (or pairs) involving $V_{Zn}$ as well as $Zn_I$ will stabilize. The isolated $V_O$s are stable up to very high temperature. The role of $O_{IS}$ in determining the electrical or optical properties in ZnO is not significant to the best of our knowledge. We feel that in this sample non-radiative recombination dominate and isolated $V_{Zn}$ complexes act as such recombination centres[23]. This perception is further confirmed from the fact that the luminescence from ZnO single crystals is greatly reduced after mechanical polishing[19]. It is well known that mechanical polishing generates subsurface $V_{Zn}$s in ZnO[32]. $V_{Zn}$ also exist in the unirradiated sample near the GBs[16]. Most probably they exist forming a different complex during growth in thermodynamic equilibrium (or with the adsorbed species). With increase of irradiation fluence individual collision cascades start overlapping and a saturation[24] of defective state is reached. Also some new kind of defect generation is expected. The antisite (Zn or oxygen antisites) defects require large energy to be formed so their concentration is very low when the system is grown in thermodynamic equilibrium. But during ion irradiation with high fluence their formation is more probable[1,4]. Formation of antisites also lower the generation rate of Zinc and oxygen vacancies and helpful to reach defect saturation. $O_{Zn}$ are acceptors and their optical transition is theoretically predicted[63] near 2.38 eV. Zinc antisites are donors but optical transition is not well known. In our opinion, major contribution of the defect level emissions at room temperature comes from $O_{Zn}$, $Zn_I$ and $V_O$ related defects. In fact, presence of at least two types of defects contributing to defect level emissions has been observed in optically detected EPR expeminets[62,65]. We have calculated the ratio of intensity at 2.43 eV and 3.27 eV and have been shown in the inset of figure 8. This ratio is thought to be an important parameter indicating the overall disorder character in the ZnO sample[20]. For the higher two fluences, the ratio shows a saturation behaviour. We have discussed possible origin of such saturation in the context of XRD and resistivity results. However, the interesting feature is that intensity of emissions at 2.43 eV and 3.27 eV both are increased due to higher two fluences. This observation is contrary to the report of Shalish et al.[64] where the UV peak is enhanced at the cost of defect related emissions. Rather, our result is close to what have been found[66] by Xiong et al. Both UV as well as defect related PL intensities have been found to increase with an average grain size within 25-75 nm. In our study, reorganization of defect structure by ion irradiation above certain fluence causes similar phenomenon in PL without significant change in grain size. So, competition of UV and defect level radiative decay cannot be a general rule in disordered ZnO. On the contrary, they may have common origin[9]. This can be felicitated from two possibilities. The UV peak in granular ZnO is defect related. The electron transition from the same defect to a deep level gives rise to sub band gap luminescence. The other possibility is that the UV peak is a real exciton transition. The restructuring of defects by ion irradiation causes increase of available carriers and the overall radiative decay, part of which is increasing defect level emissions. A temperature dependent PL investigation can only resolve problem in a definite manner.

In summary, in this study we have explored the huge possibility of low energy ion beam to create rich variety of defects in ZnO. Ar ion has been found to very much effective in producing $V_O$ defects. $V_O$s are stable at room temperature and plays important role in stabilizing $Zn_I$s. Change of colour and increase of conductivity is, most probably, initiated by the $Zn_I$-$V_O$ defect pair. Both UV and defect level PL emissions have been found to increase due to high irradiation fluence. If $V_O$s play a crucial role in inducing ferromagnetism[10,25] in ZnO, then irradiation by Ar beam should be investigated with particular emphasis. Furthermore, temperature dependent PL, temperature and depth resolved positron annihilation spectroscopy and electron paramagnetic resonance studies on low energy ion irradiated samples would be beneficial to understand "Defects in ZnO", which, theoretically and technologically, is a present day need.

Financial assistance from DST-FIST, Government of India is gratefully acknowledged. One of the authors (SC[a]) is grateful to Government of West Bengal for providing financial assistance in form of University Research Fellowship. We would like to thank Dr. Prabir Kumar Mayti, Department of Chemical Technology, University of Calcutta for helping us in performing SEM measurements.

**Figure Captions:**

FIG. 1. Electronic and Nuclear Energy Losses with penetration Depth of 1.2 MeV $Ar^{+8}$ ions in ZnO as calculated from SRIM code.

FIG. 2. Variation of Ionization, Oxygen Vacancy and Zinc Vacancy with ion range as calculated from SRIM code.

FIG. 3. Colour of (a) unirradiated and irradiated with 1.2 MeV $Ar^{+8}$ ions of fluence (b) $1 \times 10^{15}$, (c) $5 \times 10^{15}$ and (d) $5 \times 10^{16}$ ions/cm$^2$ ZnO samples.

FIG. 4. SEM pictures of (a) unirradiated and (b) irradiated with 1.2 MeV $Ar^{+8}$ ions of $5 \times 10^{16}$ ions/cm$^2$ ZnO samples.

FIG. 5. XRD spectra of (a) unirradiated and irradiated with 1.2 MeV $Ar^{+8}$ ions of fluence (b) $1 \times 10^{15}$, (c) $5 \times 10^{15}$ and (d) $5 \times 10^{16}$ ions/cm$^2$ ZnO samples.

FIG. 6. Enlarged view of the (101) peak region of the ZnO XRD spectra in the range of 2θ from 36º to 37.5º. Inset: Similar view of the (002) peak of the ZnO XRD spectra in the range of 2θ from 34.4º to 35.5º.

FIG. 7. Variation of FWHM of (101) peak (black square), Ratio of intensity (101) to (002) peak (blue down triangle), and (101) Peak Position (red circle) with Irradiation Fluence.

FIG. 8. PL Spectra of (a) unirradiated and irradiated with 1.2 MeV $Ar^{+8}$ ions of fluence (b) $1 \times 10^{15}$, (c) $5 \times 10^{15}$ and (d) $5 \times 10^{16}$ ions/cm$^2$ ZnO samples. Inset: Variation of PL intensity ratio of peak at 3.27 eV to peak at 2.43 eV with irradiation fluence.

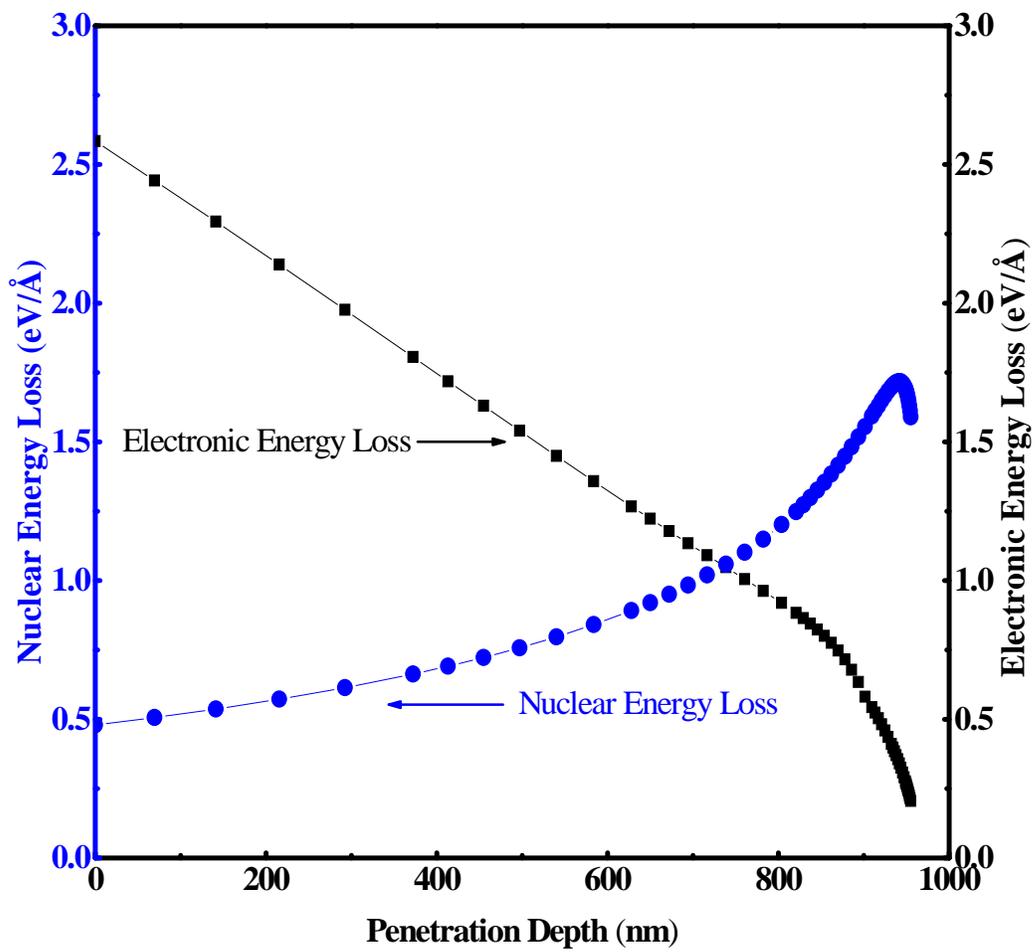

**FIG. 1.**

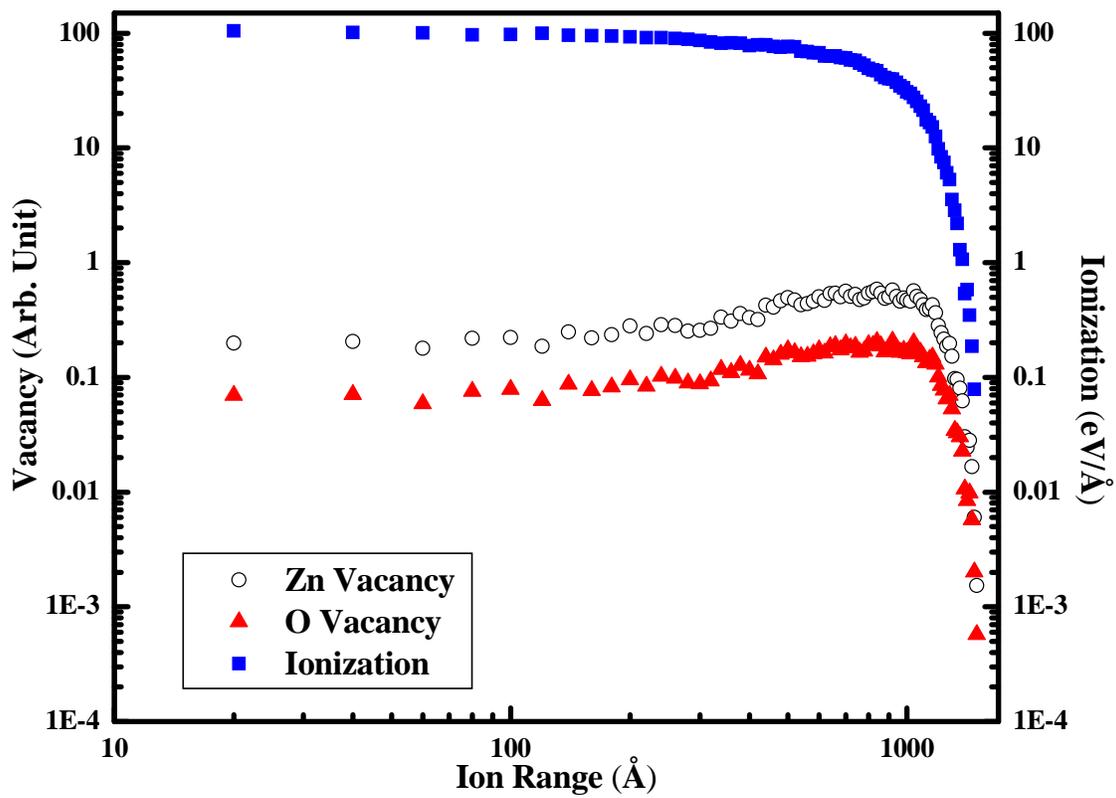

**FIG. 2.**

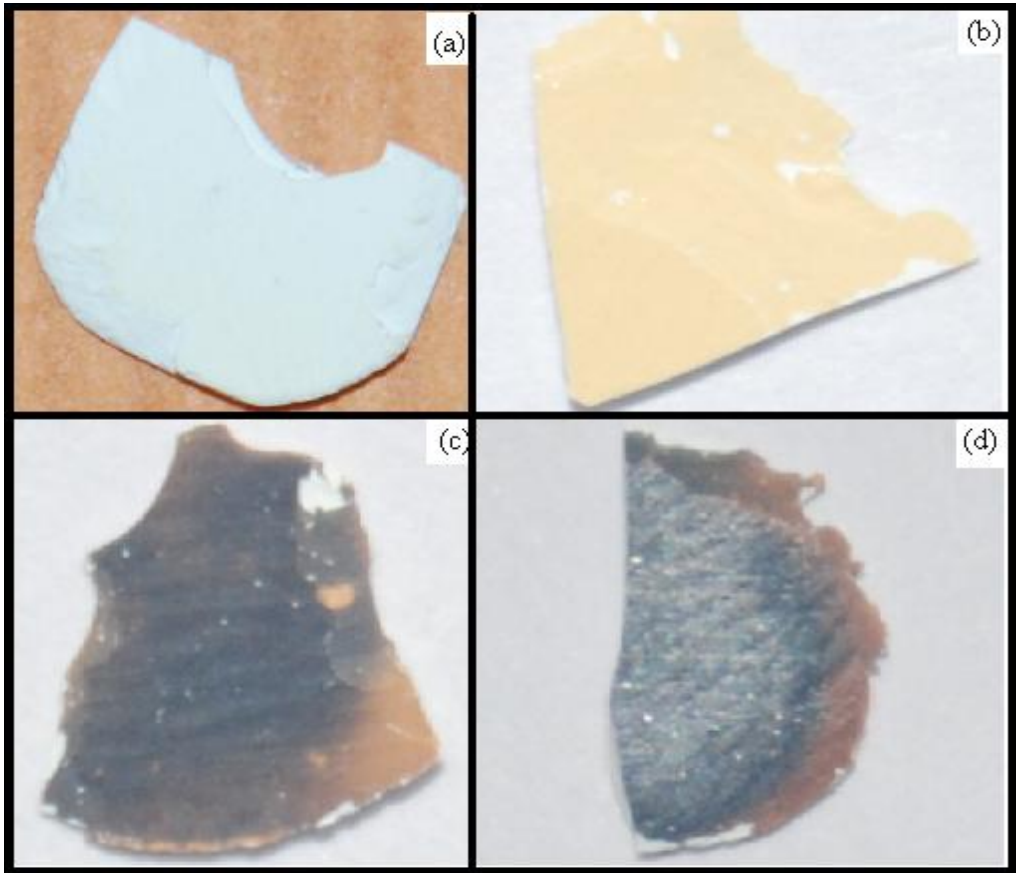

**FIG. 3.**

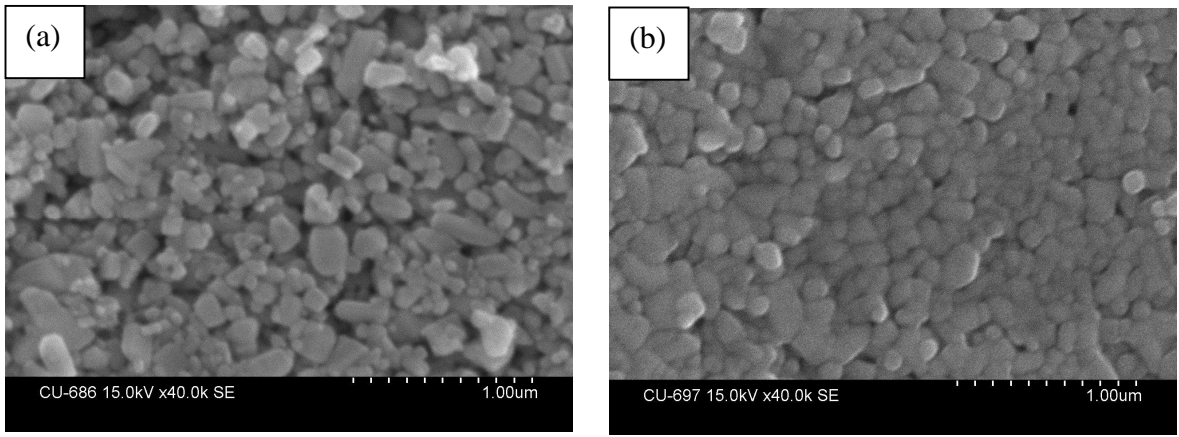

**FIG. 4.**

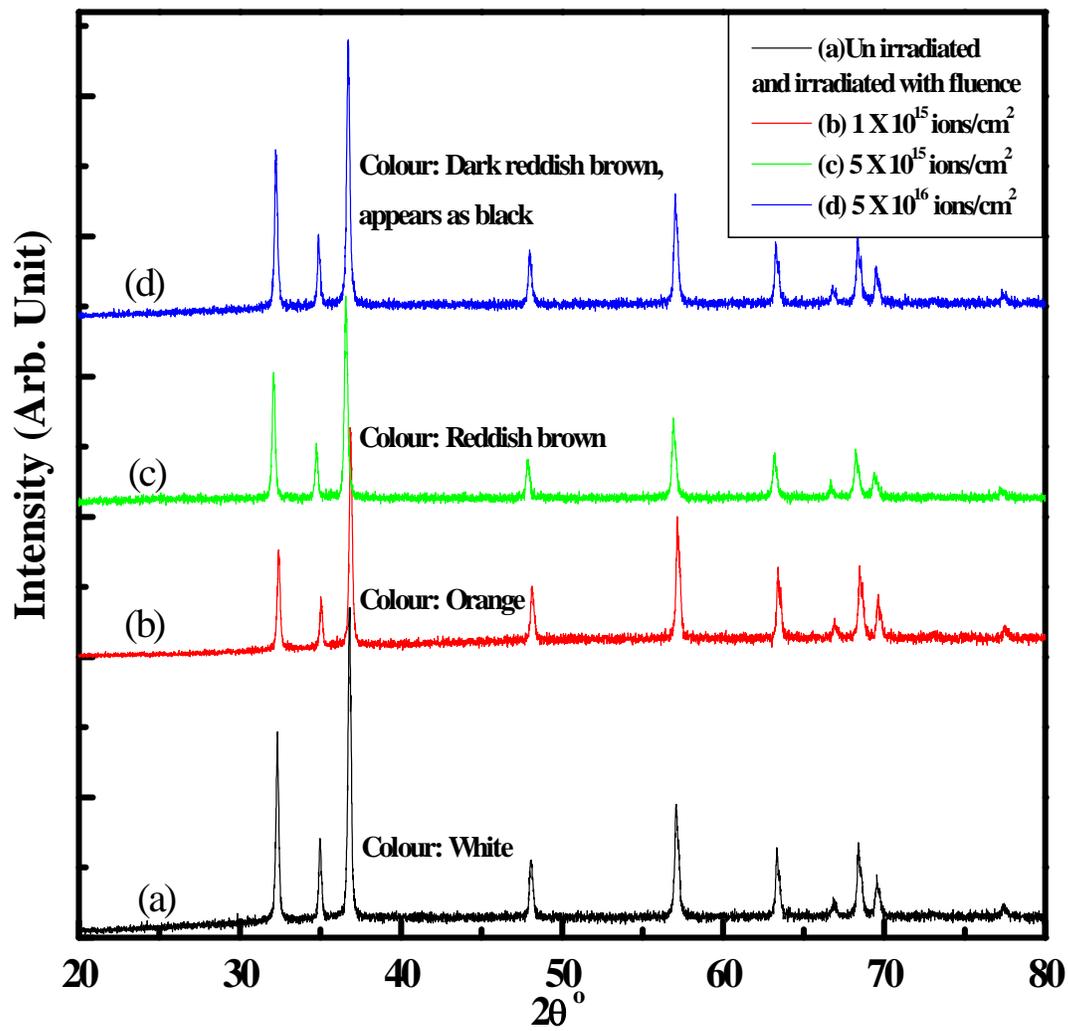

**FIG. 5.**

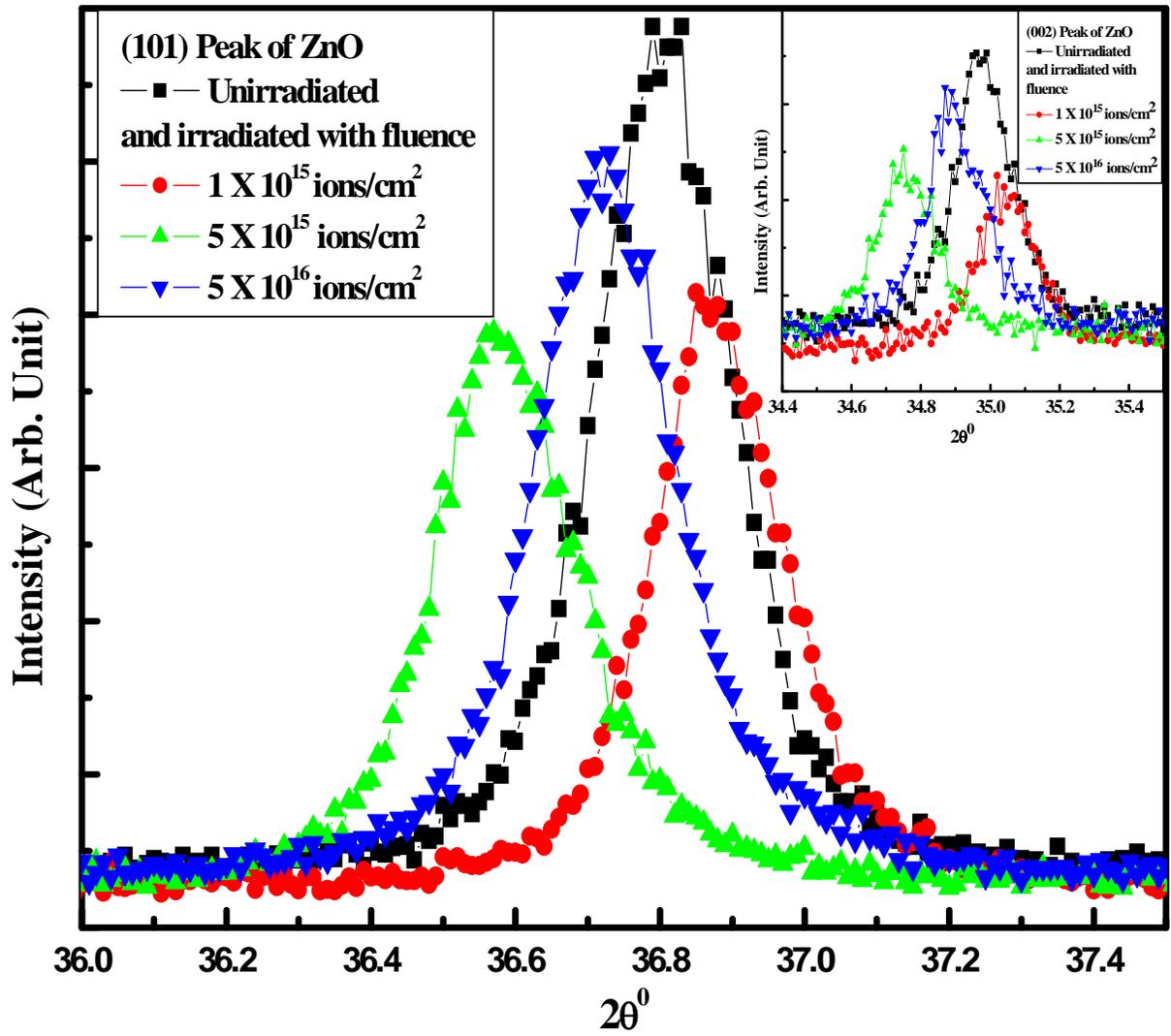

**FIG. 6.**

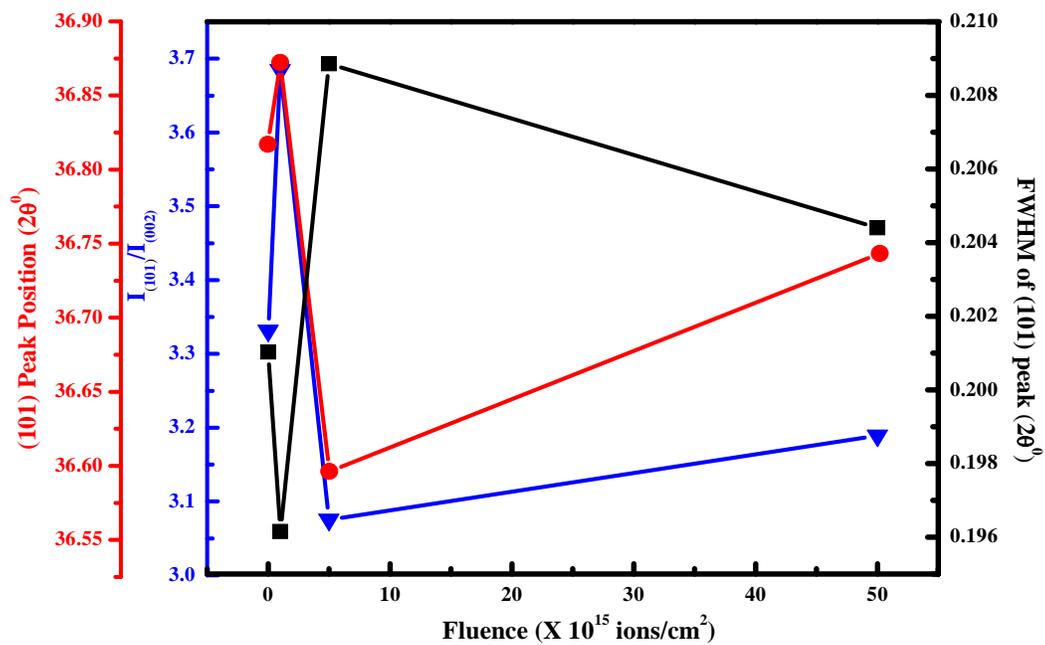

**FIG. 7.**

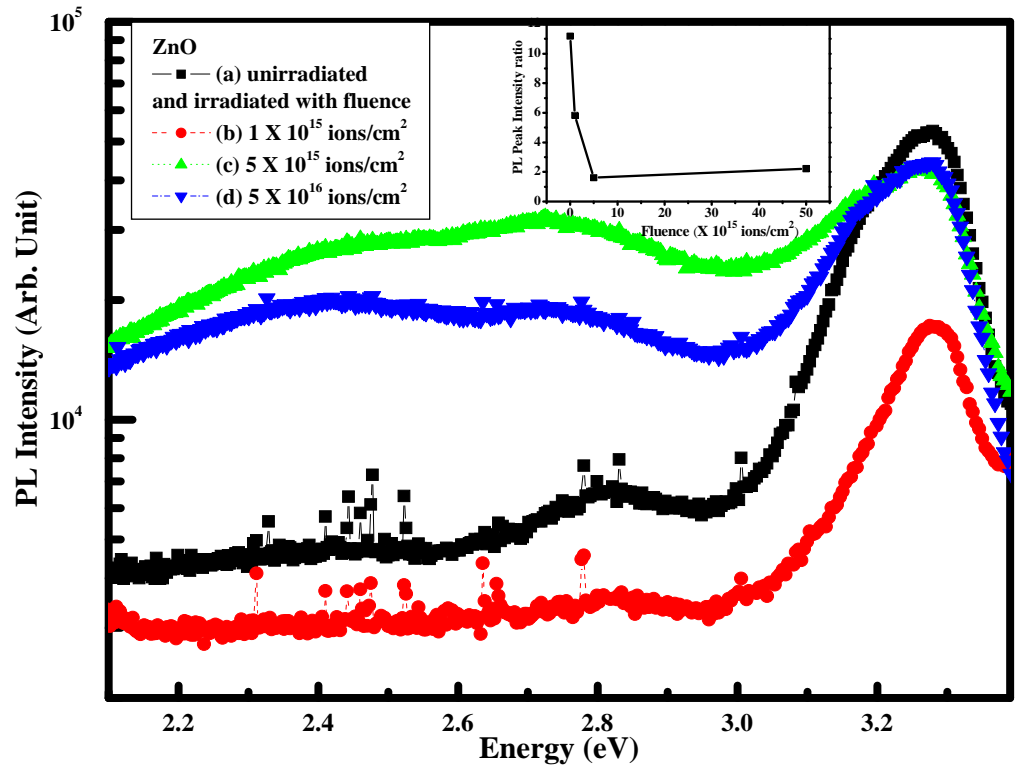

**FIG. 8.**